\documentclass[epsf,useAMS,fleqn]{mn2e}
\pdfoutput=1
\usepackage{graphicx}
\usepackage{times}
\usepackage{epsfig}
\usepackage[english]{babel}
\usepackage{listings}
\usepackage{color}
\usepackage{amsmath}
\usepackage{epstopdf}
\usepackage{url}

\usepackage{fixme}
\usepackage{subfig}
\usepackage{booktabs}
\usepackage{threeparttable}
\usepackage[colorlinks]{hyperref}
\renewcommand{\eqref}[1]{Eq.~\ref{#1}}

\title[Moving Windowed Power Search]{A new method to detect solar-like oscillations at very low S/N using statistical significance testing}
\author[M. N. Lund et al.]{Mikkel~N.~Lund$^{1,2,3}$\thanks{E-mail: mikkelnl@phys.au.dk}, William~J.~Chaplin$^{2,3}$\thanks{E-mail: w.j.chaplin@bham.ac.uk}, Hans~Kjeldsen$^{1,3}$\thanks{E-mail: hans@phys.au.dk}\\
$^1$ Department of Physics and Astronomy, Aarhus University, DK-8000 Aarhus C, Denmark\\
$^2$ School of Physics and Astronomy, University of Birmingham, Edgbaston, Birmingham B15 2TT, UK\\
$^3$ Stellar Astrophysics Centre (SAC), The Danish National Research Foundation}

\begin{document}

\maketitle

\begin{abstract}

We introduce a new method to detect solar-like oscillations in
frequency power spectra of stellar observations, under conditions of
very low signal to noise. The Moving-Windowed-Power-Search, or MWPS,
searches the power spectrum for signatures of excess power, over and above slowly varying (in frequency) background contributions from
stellar granulation and shot or instrumental noise. We adopt a
false-alarm approach (Chaplin et al. 2011) to ascertain whether flagged excess power, which
is consistent with the excess expected from solar-like oscillations,
is hard to explain by chance alone (and hence a candidate detection).

We apply the method to solar photometry data, whose quality was
systematically degraded to test the performance of the MWPS at low
signal-to-noise ratios. We also compare the performance of the MWPS
against the frequently applied power-spectrum-of-power-spectrum
(PS$\otimes$PS) detection method. The MWPS is found to outperform the
PS$\otimes$PS method.

\end{abstract}


\section{\textbf{Introduction}}
\label{sec:intro}

Asteroseismology on solar-like oscillators can provide important
knowledge on the internal structure and evolution of these stars, and
in the case of planetary systems this knowledge is needed for much of
the analysis on the planets themselves. The analysis and
interpretation of the stellar oscillations relies of course on the
actual detection of the signatures of the solar-like
oscillations. These signatures manifest as a pattern of near regularly
spaced peaks, and hence a power excess, in the frequency spectrum of
the stellar time series. The advent of space-based photometric
observations, such as those provided by the NASA \textit{Kepler}
Mission (Borucki et al. 2010; Koch et al. 2010), is providing large
quantities of exquisite-quality data for asteroseismology (Gilliland
et al. 2010). The first stage of asteroseismic analysis of solar-type
stars often now involves the application of automated analysis
pipelines to search for signatures of stellar oscillations. These
signatures may then be analysed in detail if a positive detection is
obtained.

Many of the methods exploit the near regularity (or periodicity) in
frequency of the oscillation frequencies, for example from computation
of the autocorrelation of the timeseries (equivalently the power
spectrum of the power spectrum, or PS$\otimes$PS) (see, e.g., Hekker et
al. 2010; Mosser \& Appourchaux 2009), or the autocorrelation of the
power spectrum (see, e.g., Campante et al. 2010; Karoff et al. 2010) and
variants thereof (see, e.g., Verner \& Roxburgh 2011; Huber et
al. 2009; Mathur et al. 2010). These methods will have difficulty in detecting a signal if
the regular structure is obscured, as for example in stars with many
mixed-modes, or in stars where the stellar oscillation signal is
buried in noise so that beating with the noise becomes significant.

Here, we present a method that instead searches just for the power
excess originating from stellar oscillations. This
\emph{Moving-Windowed-Power-Search} method (hereafter MWPS) uses a
moving window in frequency, and a statistical significance test is
employed to see if any excess power flagged over and above the
expected sources of background noise could be caused solely by
fluctuations of the noise. The method does not rely on detecting any
regular frequency structure in the stellar oscillation spectrum and is
shown to work extremely well under low signal to noise (low SNR)
conditions.
To see how the MWPS method performs compared to currently used methods, we test it against the PS$\otimes$PS method, which is one of the most commonly used techniques for extracting the large frequency separation. Both methods are tested on solar data, to which increasing amounts of white noise were added in the time series.

The layout of the rest of the paper is as follows. In Section
\ref{sec:mwps} we introduce the principles of the MWPS method, and
discuss in detail its implementation. The PS$\otimes$PS, with which we
compare the MWPS, is discussed in Section \ref{sec:comp}. We present
results of testing both methods on solar data in Section
\ref{sec:test}. Concluding remarks are given in Section
\ref{sec:conc}.

\section{Moving-Windowed-Power-Search}
\label{sec:mwps}

 \subsection{Overview}
 \label{sec:overview}


The oscillation spectrum of a solar-type star consists of a series of regularly spaced peaks with heights modulated by a broad envelope. This envelope of power due to oscillations may be approximated by a Gaussian, with
$\nu_{\rm max}$ defined as the frequency at the peak of the power
envelope of the oscillations where the observed modes present their
strongest amplitudes (Chaplin et al. 2011). 

The power\footnote{The power spectrum was calculated using a sine-wave fitting method (see, e.g., Kjeldsen 1992, Frandsen et al. 1995) which is normalized according to the amplitude-scaled version of Parseval's theorem (see, Kjeldsen \& Frandsen 1992), in which a sine wave of peak amplitude, A, will have a corresponding peak in the power spectrum of $\rm A^2$.} 
due to the oscillations sits on a slowly varying (in frequency) background power
that we assume is dominated by contributions from shot/instrumental
noise, stellar granulation and activity.  When the observed power in
the oscillations relative to the background is high (i.e., high SNR),
the power excess due to the oscillations will be clearly
visible. However, at low SNR, statistical fluctuations in the
background power may swamp the oscillations signal, so that the excess
is much harder to see.

The MWPS method searches the spectrum for the presence of excess
power, over and above the slowly varying background. A statistical
false-alarm (null hypothesis) test is applied to ascertain whether any
flagged excess is hard to explain by chance alone, i.e., random
fluctuations of the background. Such cases may be flagged as candidate
detections, subject to some \emph{a priori} constraints regarding the
total excess power solar-like oscillations would be expected to show
in the flagged frequency range (of which more in
Section~\ref{sec:width} below).

The method is clearly reliant on providing an accurate estimate of the
underlying background, since any excess is defined \emph{relative to}
that background. The background fitting is discussed in detail below,
in Section~\ref{sec:bg}. First, we discuss the statistics of the MWPS
(Section~\ref{sec:how}), and logistics relating to how the frequencies
are searched for evidence of a power excess (Section~\ref{sec:width}).

\subsection{\textbf{Statistics of the MWPS}}
\label{sec:how}

At each tested frequency we estimate a power-to-background ratio,
$\text{PBR}_{\rm tot}$, over a pre-defined window in frequency. This ratio
is defined according to
 \begin{equation}
 \rm PBR_{\rm tot}=\frac{P_{\rm tot}}{B_{\rm tot}},
 \end{equation} 
where $\text{P}_{\rm tot}$ is the total power across the test
frequency window, i.e., it includes any excess power above the
underlying fitted background, and the contribution from the
background; and $\text{B}_{\rm tot}$ is the total power in the
background, i.e., the area below the fitted background
curve\footnote{$\text{P}_{\rm tot}$ and $\text{B}_{\rm tot}$ was found
  by the sum of the discrete values}. Note that this definition of
PBR$_{\rm tot}$ corresponds to the value of SNR$_{\rm tot}$+1 in
Chaplin et al. 2011.  The value of $\text{PBR}_{\rm tot}$ is
normalised in such a way that it will tend to $\text{PBR}_{\rm
  tot}\sim 1$ in the case of no excess power from oscillations.
The motivation for using a window instead of simply the entire power spectrum is, that by using a frequency range much larger than the one containing the excess power from oscillatons, the value of $\rm PBR_{\rm tot}$ would be much degraded, reducing the chances of making a detection. 
The tested frequencies and associated window widths are discussed in
Section~\ref{sec:width} below.

We assume that the individual bins of the power spectrum follow
$\chi^2_2$ statistics, i.e. that they follow a chi-squared
distribution with two degrees-of-freedom (d.o.f.). As the sum of two
such independent distributions $\chi^2_k + \chi^2_m$ is yet another
$\chi^2_n$ distribution with $n=k+m$ d.o.f., the sum of power from the
$N$ independent\footnote{No oversampling is used in the computation of
  the power spectrum} bins included in the envelope interval will then
be distributed according to a $\chi^2$ distribution with
$2N$-d.o.f. We may therefore compare the value of $\text{PBR}_{\rm
  tot}$ with a $\chi^2$ 2$N$-d.o.f. statistic (Appourchaux et al. 2004). The functional form of
the $\chi^2$ distribution is given by:
 \begin{equation}
 f(x;k)=\frac{1}{2^{k/2} \Gamma(k/2)} x^{k/2 -1} e^{-x/2},
 \end{equation}
where $k$ is the number of d.o.f., $\Gamma$ is the gamma function and
$x$ is the calculated value of $\text{PBR}_{\rm tot}$ (actually $x=\text{PBR}_{\rm tot} k$).

Knowing the distribution, it is possible to calculate a corresponding
probability, or $p_{\rm value}$, for the obtained value of
$\text{PBR}_{\rm tot}$, i.e. the probability of observing a value $\ge
x$ by chance, assuming the \emph{null} hypothesis is true, i.e.,
 \begin{align}
 p_{\rm value} &= 1 - F(x;k)\label{eq:sur}\\
 &= 1 - \int_x^{\infty}f(x;k) dx\nonumber
 \end{align}
where $F(x;k)$ is the cumulative $\chi^2_k$ distribution.  Needless to
say, the lower the $p_{\rm value}$ the less likely it is that the
observed power can be attributed to random fluctuations of the
underlying background. If the $p_{\rm value}$ falls below a
pre-defined \emph{significance level} (or \emph{false alarm
  probability}), $\alpha$, the \emph{null} hypothesis can be rejected
and the observed power is then assumed to be statistically
significant. Here, we have set $\alpha = 0.01$.
 
The threshold value of $\text{PBR}_{\rm tot}$ required to reject the
null hypothesis is given by:
 \begin{equation}
 P(\rm PBR_{\rm tot} \geq \rm PBR_{\rm threshold}) = \alpha,
 \end{equation}
which amounts to finding the PBR at which $p_{\rm value} = \alpha$.  In
sum, the "false-alarm" approach means that every calculated $p_{\rm
  value}$ falling below $\alpha$ is flagged as marking the presence of
significant excess power, which could be attributable to stellar
oscillations.  The procedure thereby gives us a probability curve
across the frequency interval for each of the selected windows.

 \subsection{\textbf{Tested frequencies and window widths}}
 \label{sec:width}

We make no \emph{a priori} assumptions on the true value of
$\nu_{\rm max}$ and thereby on the range of frequencies covered by any
Gaussian power envelope that might be present. However, we do allow
for the fact that the FWHM in frequency of the oscillation envelope
is known to vary to good approximation as $\nu_{\rm max}/2$ (see,
e.g., Stello et al. 2007; Chaplin et al. 2011), and so test for the
presence of excess power over windows of the spectrum whose width in
frequency is varied accordingly.

The centre frequency of each test window may be regarded as a proxy
for $\nu_{\rm max}$. When the MWPS is applied at low frequencies the
width of the test window should be set fairly narrow, since any
oscillation signal present will also be confined over a narrow range
of frequencies. A wide test window -- i.e., one significantly wider
than the oscillation power envelope -- would reduce the underlying
PBR, and hence reduce the chances of making a detection with the
MWPS. The same logic dictates that the window width must be set much
wider when the test is applied at higher frequencies (too narrow a
window will result in signal being missed). 

Assuming a Gaussian-like power excess, there is notionally an optimum
window width that will maximize the underlying PBR, which corresponds approximately
to $1.17$-times\footnote{Estimated from tests on toy models of the power spectrum with varying values of e.g. the shot noise and the parameters descibing the Gaussian power envelope} the FWHM of the power envelope, which is $0.59\times\nu_{\rm
  max}$, i.e., $0.59$-times the test frequency. This width may not always
yield optimum results in practice, due to the importance at low SNR of
beating with background noise, and the impact of realization noise
from any oscillations that are present. We make suitable allowance for
this in our adopted search strategy.


We begin by selecting 40 frequencies spread evenly in the range from
1000 to $5000\,\rm \mu Hz$ (adopted range for solar-like oscillations). An optimum window width is calculated for
each of these $\nu_{\rm max}$ proxy frequencies, giving a total of 40 windows with which to
test the power spectrum. Each window is then moved through the power
spectrum in steps set by a pre-defined lag frequency (overlaps, but not gaps, between frequency ranges encompassed by the respective windows are allowed), and the MWPS test is applied at each location. The check for excess power at each
location involves first calculating the total power in the envelope
interval set by the window width, which may then be compared to the
total power found below the fitted background (see
Section~\ref{sec:bg} below for details of the background fitting).

We do not vary the window size as each window is moved through the
spectrum. The reason for this is twofold: First, it turns out that
varying the window width tends to skew the resulting SNR and $1/p_{\rm
  value}$ curves, mainly due to overlap effects as the window grows in
size towards higher frequencies, and this inevitably introduces an
unwanted offset in the peak position of the detectability, i.e., in
the estimated $\nu_{\rm max}$. Second, use of a single varying window
does not take into account that the selected (notionally optimal)
window size may not be optimal in practice, as noted above.

Application of the MWPS as outlined above yields an SNR (or
equivalently PBR) curve and a corresponding $1/p_{\rm value}$ curve,
as a function of frequency for each of the adopted 40 window widths.
An ensemble analysis of those curves which have values of $1/p_{\rm
  value}$ that exceed the detection threshold then yields an estimate
of $\nu_{\rm max}$. We discuss this procedure in detail in
Section~\ref{sec:test} below.

We then apply a sanity check on the results.  The total power in the
oscillations spectrum is a strong function of $\nu_{\rm max}$, i.e.,
the higher $\nu_{\rm max}$, the smaller is the expected total
oscillations power. We apply the following formulae from Chaplin et al. (2011)
to estimate the expected total power, based on the estimated $\nu_{\rm
  max}$:
\begin{align}
&\rm P_{tot} \approx 1.55 A_{max}^2 \eta^2\frac{\nu_{max}}{\Delta\nu}\\
&\eta^2 = sinc\left[\pi/2 \left(\frac{\nu_{\rm max}}{\nu_{Ny}}\right) \right]
\end{align}

Here $\rm P_{tot}$ is the total power underneath the Gaussian shaped envelope in the range $\pm \rm \nu_{max}/2$ around the estimated $\nu_{\rm max}$ value, and $\rm A_{max}$ is the radial-mode amplitude at $\nu_{\rm max}$. The factor $\eta^2$ is a correction for the apodization of the oscillation signal the closer $\nu_{\rm max}$ is to the Nyquist frequency, $\nu_{Ny}$, and $\Delta\nu$ is the average large separation (see discription to Eq.~\ref{eq:split}) which may be estimated from $\nu_{\rm max}$ (see Eq.~\ref{eq:scaling}).  
This prediction may then be compared to the measured excess
power in the spectrum. If the measured excess is significantly higher
than the predicted excess (i.e., by some multiple of the combined
uncertainties on the prediction and measurement) the candidate
detection is flagged as questionable. Strong, narrow-band artefacts
can produce signatures of this type, hence knowledge of known
instrumental issues is important to help try to verify the robustness
(or otherwise) of any claimed detection.

 \subsection{\textbf{Fitting the background}}
 \label{sec:bg}
 
Since the success of the MWPS method relies heavily on a good fit of
the background, we devote this section to a description of the
background fitting process.
 
We took the approach of a $\chi^2$-minimization instead of directly
maximising the likelihood function of the power spectrum (having a
$\chi^2_2$-statistic) mainly to decrease the computation time for the
fitting.  In order to use a $\chi^2$ procedure the power spectrum is
binned linearly in $\text{ln}(\rm \nu)$. This ensures first of all
that the points used in the background fit are independent and
secondly, that the density of points in frequency is larger towards
lower frequencies where the bulk of variations due to background
phenomena occur. The averaged datum with the smallest number of binned
frequencies used in the fitting tests (Section~\ref{sec:test}) came
from an average made over $\sim550$ frequency bins, which makes it
safe to assume that the points will have a normal
distribution,\footnote{With $k>50$ a normal distribution can be
  assumed (e.g. Box, Hunter and Hunter 2005)} justifying the use of a
$\chi^2$-minimisation scheme:
 \begin{align}
 \chi^2 &= \sum_{i=1}^N \left[ \frac{y_i - bg(x_i;\textbf{a})}{\sigma_i} \right]^2,\\
 \sigma_i &= \frac{bg(x_i;\textbf{a})}{\sqrt{n}}.
 \label{eq:sumech}
 \end{align}
Here the sum is over the binned data, $y_i$ is the value of the binned
datum $i$. $bg(x_i;\textbf{a})$ is the value of the fitted background function,
where $\textbf{a}$ are the dependent variables for the background function. The
weighting, given by $\sigma_i$, depends as seen on the number of
frequencies, $n$, included in the binning.
  
The function describing the background signal consists of power laws
(Harvey 1985), each of which describes a specific physical
phenomenon. The power laws included describe the signal due to,
respectively, stellar activity and granulation, given by (see,
e.g., Hekker et al. 2010; Mathur et al. 2011):
 \begin{equation}
 bg(x_i;\textbf{a}) = \frac{A_g}{1+(B_g x_i)^c} + \frac{A_a}{1+(B_a x_i)^2} + S_N.
 \label{eq:bg}
 \end{equation}
In this equation $A_g$ and $A_a$ give the power of granulation and
activity respectively, while $B_g$ and $B_a$ represent the
characteristic time scales for the two phenomena as the power remains
approximately constant on time scales longer than $B_{g/a}$ (or
equivalently frequencies lower than $B_{g/a}^{-1}$), and drops off for
shorter time scales. In the fitting, the exponent, $c$, of the granulation slope is kept as a free
parameter, while the exponent for the activity slope is fixed to a value of 2. The fixed value of 2 for the slope of the activity component arises from the fact that we assume an exponential decay of the activity regions as a function of time\footnote{The fourier transform of an exponentially decaying function is a Lorentzian function, i.e. a power law as in Eq.~\ref{eq:bg} with an exponent of 2}.
The constant parameter $S_N$ gives the photon shot noise. Whilst the method was tested we omitted binned data from the first $250\rm\ \mu Hz$ of the power spectrum, so that only the granulation and the shot noise was needed to fit the background. The reason for only including these two components is that phenomena related to activity, such as e.g. spots (both intrinsic variations and rotational modulation), mesogranulation and supergranulation all have characteristic time scales longer than $4000$ s (see, e.g., Harvey 1985), equivalent to frequencies below  $250\rm\ \mu Hz$.


\begin{figure}
\centering
\includegraphics[trim=0.6cm 0.4cm 1.75cm 1.3cm, clip=true, width=1\columnwidth]{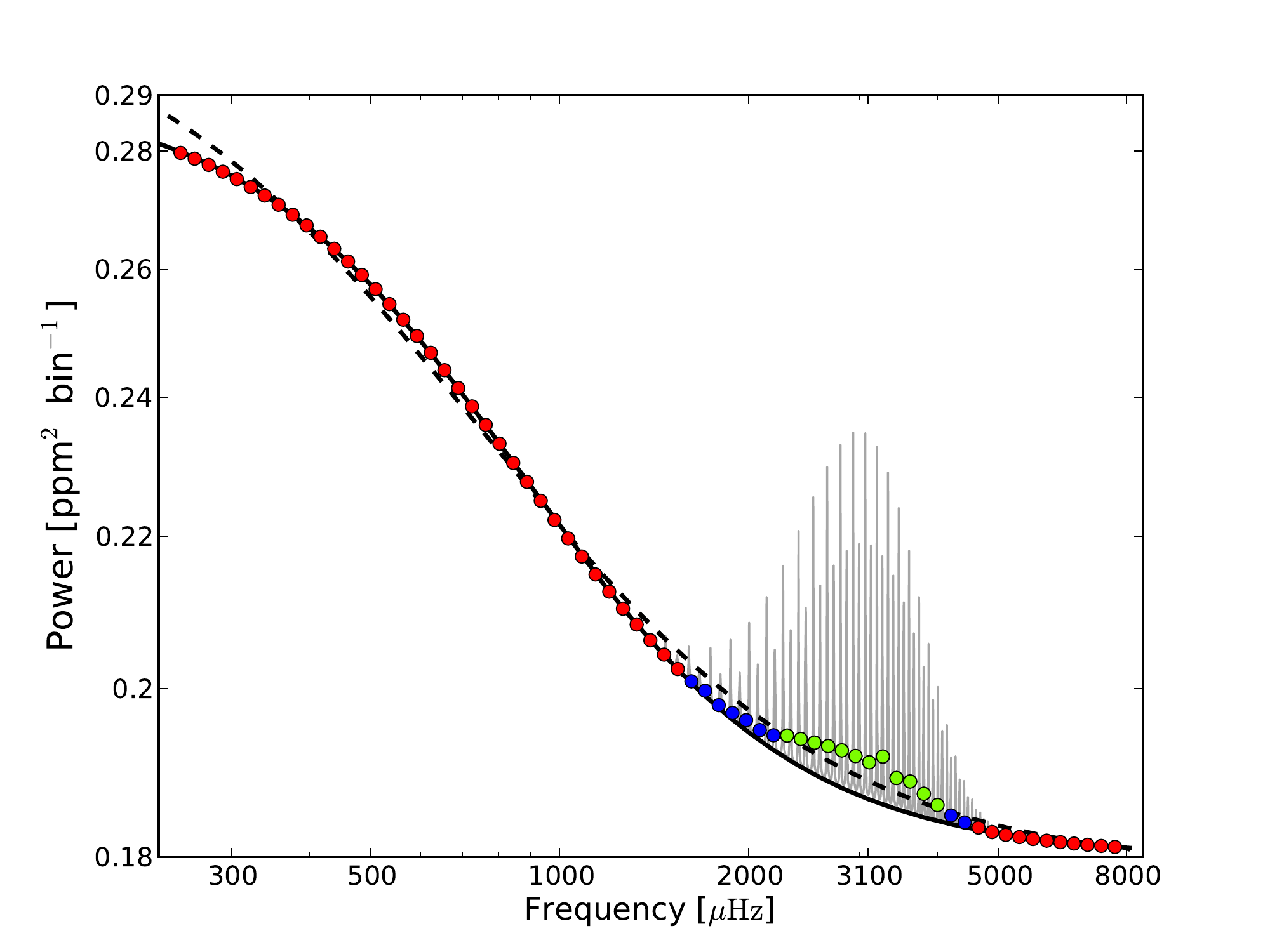}
\footnotesize

\caption{Illustration of the impact of applying the runs-test, here
  shown on artificial noiseless data shown as the limit power spectrum in grey. This limit spectrum is modelled as a series of Lorentzian peaks with a regular frequency spacing and a height modulated by a Gaussian envelope (see Eq.~\ref{eq:split}), which sits on top of a slowly varying background function. The points in the figure are the
  values for the binned spectrum. The dashed black curve shows the
  background fit to all the binned points, and as seen the fit is
  somewhat offset from the expected level. The green points mark the
  longest consecutive series of points above the background curve. The
  solid black curve gives the best fit to the red points, i.e., after
  removing the points flagged by the runs-test (green) along with the
  additional points on either side, given by the blue points.}

\label{fig:offset} 
\end{figure}


A general problem when fitting the background to the power spectrum is
of course that any excess power from stellar oscillations will offset
the background fit, primarily through an overestimated value of the
constant shot noise component, and make it very difficult to pin down
the correct value for the slope of the background function (see
Fig.~\ref{fig:offset}) which again could produce unwanted peaks in the
calculated SNR curves.  As this offset will lower the PBR value it can
have a determining impact on the performance of the MWPS method,
especially in high noise cases.

To boost the robustness of our method and circumvent the offset from a
potential power excess, a Wald-Wolfowitz runs-test (see Barlow 1989)
is applied to the intial background fit to check for non-random
offsets around the fitted curve. This type of information in not
obtainable from the fitting procedure as the $\chi^2$-minimization is,
by its very nature, completely blind to the sign of any deviation,
whereas the Wald-Wolfowitz runs-test is blind to anything but the sign
of the offset making it a good complimentary test.  Any power excess
which is strong enough to offset the background fit will be
distributed in such a way that there will be a relatively long run
(i.e., a consecutive series) of points above the fitted curve at the
position of the power excess (see Fig.~\ref{fig:offset}). If the
false-alarm probability obtained from the runs-test falls below a
given value -- indicating that the chances of the run occuring by
chance are low -- it may be assumed that the background fit is offset
due to some power excess not described by the background function.

The points comprising the longest consecutive series above (in power)
the background fit are removed in addition to a certain user defined
number of points at the low- and high-frequency ends of this
series\footnote{We set the constraint that the series must have a
  length of at least three points}, and the background is fitted anew
to the remaining binned points.  In our testing we generally removed
seven additional points at the low-frequency end, and three at the
high-frequency end (remember that as the binned points are linear in
$\text{ln}(\rm \nu)$ the distance in frequency between two points at
the low-frequency end is much shorter than the equivalent distance at
the high-frequency end).
  
If the best-fitting $\chi^2$-value remains high after this procedure has been
applied, an extra power law could potentially be included in the
background fit until a satisfactory fit is obtained. A good fit
corresponds to a reduced $\chi^2$-value of $\chi^2_r \sim 1$. If the
reduced $\chi^2$-value is lowered after the removal of the longest run
from the runs-test, the MWPS procedure continues, even if this removal
of points worsens the result from the runs-test. In this sense the
probabilities from the runs-test are used more as indicators of a good
fit, while the configuration of runs is actively used.
\begin{figure*}
 \centering
 \includegraphics[width=1.5\columnwidth]{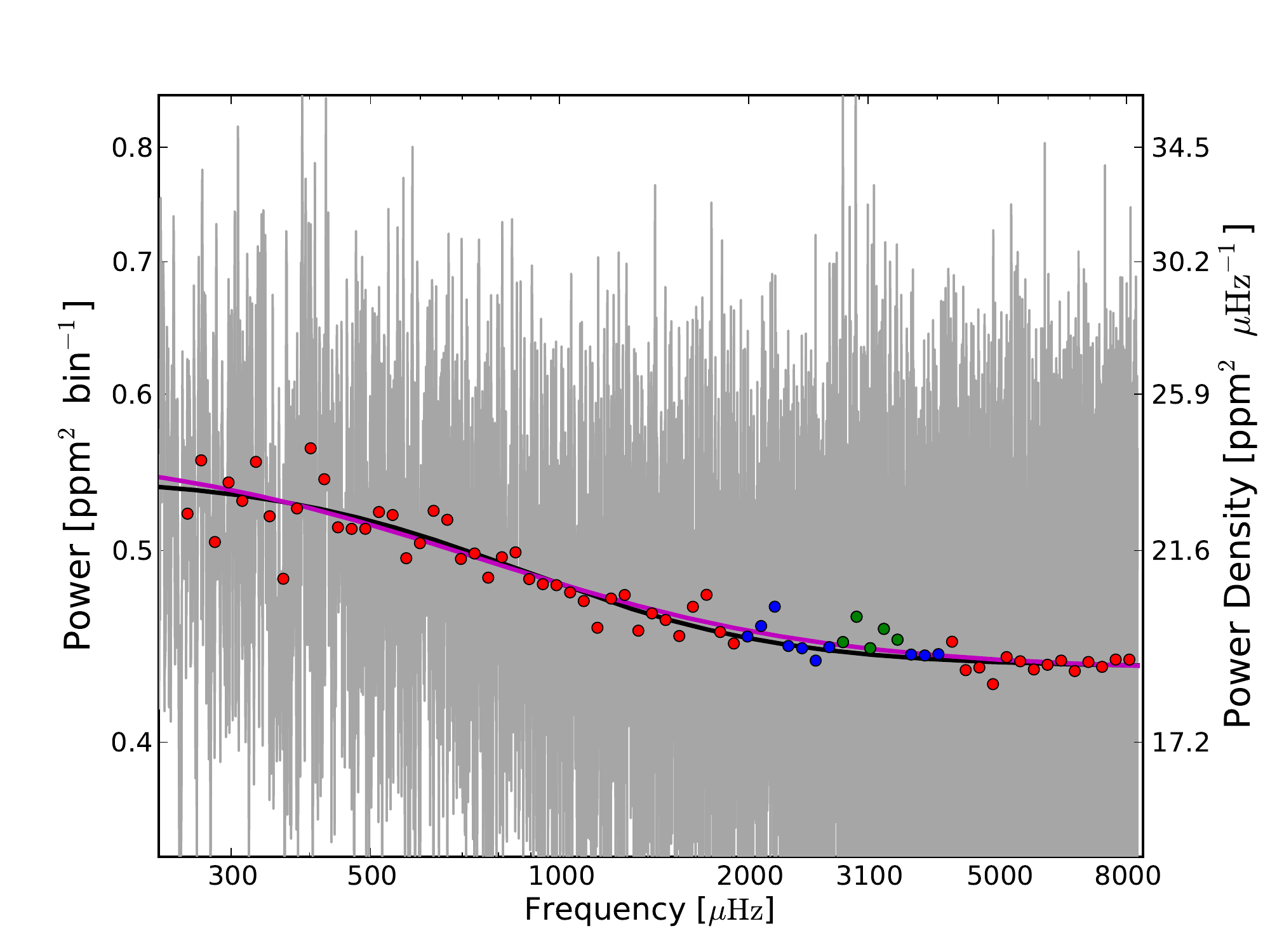}
 \footnotesize
 \caption{Example background fit for one of the $\sigma=280$ ppm
   realizations. The $1\rm\ \mu Hz$ smoothed power spectrum is plotted
   in gray, from 230 to $8500\rm\ \mu Hz$. The red points show the
   binned power spectrum. The longest run found by the runs-test is
   shown with the green points, and the additional points to be
   removed are plotted in blue. Also shown is the fitted background
   function before (magenta) and after the exclusion of the green and
   blue points (black).}
 \label{fig:DJ} 
\end{figure*}
We employed a Nelder-Mead simplex\footnote{This algorithm along with
  the rest of the code is written in the programming languages
  \textsc{python} (\url{http://www.python.org}) and \textsc{cython}
  (\url{http://cython.org})} algorithm for the minimisation of the
$\chi^2$-value, as this seems to perform in a very robust manner albeit at
the expense of being somewhat slower than other frequently applied
methods, e.g. steepest descent methods.

The starting values of the parameters of the background model were found as follows: For the granulation parameters -- i.e., the granulation time scale $B_g$ and amplitude $A_g$ -- we use information from the first derivative of the binned power spectrum in logarithmic units. This derivative is found by applying a third-order Savitzky-Golay\footnote{In a Savitzky-Golay filter one makes for every data point, $(x_i, f_i)$, a least-squares fit of a polynomial function (of a given order) to the data points within a window around this datum. The resulting smoothed value at $x_i$, is then the value of the polynomial at this position. In this sense the smoothing is performed in a moving window} (Savitzky \& Golay 1964) smoothing filter to the binned power spectrum, from which the first derivative is directly obtainable. For the Savitzky-Golay filter we chose a window width of 41 points.
In a first derivative curve like the one found, peaks will
correspond to a rise or fall (i.e., a change) in power spectral
density as a function of frequency. As mentioned in connection to
Eq.~\ref{eq:bg}, the contribution from granulation in the power
spectrum falls at frequencies above $B_g^{-1}$. The granulation time
scale is expected to be shorter than the equivalent time scale for the
activity and so we identify the highest frequency peak in the first
derivative curve as originating from the change in power due to the
granulation, from which one can readily find the associated time
scale. For the amplitude $A_g$ we find, in the first derivative curve,
the frequency of the first dip to the low-frequency side of the
granulation timescale peak, and use the power spectral density in the
power spectrum at this frequency as the estimate for $A_g$.  The shot
noise level is estimated from the mean power spectral density at the
high-frequency end of the power spectrum.  Even though the Nelder-Mead
algorithm is intended for localised optimisation, we found no
significant change in the resulting background fits when the starting
values for the optimisation were varied slightly, which is to be
expected given the low number of fitting parameters.

\section{\textbf{Comparison method}}
\label{sec:comp}

We have compared the MWPS method against the
power-spectrum-of-power-spectrum, or PS$\otimes$PS, method. For
solar-like oscillators the PS$\otimes$PS method works well because
high-order p-modes follow to good approximation a regular, periodic
pattern in frequency (Tassoul 1980), described by
 \begin{equation}
 \nu_{n,\ell} = \Delta\nu (n + \frac{1}{2}\ell + \epsilon) - \ell(\ell+1)D,
 \label{eq:split}
 \end{equation}
where $n$ is the radial order and $\ell$ is the angular degree. $\Delta\nu$
is the so-called large separation, corresponding to the spacing in frequency between consecutive radial overtones, $n$, having the same angular degree, $\ell$. $\epsilon$ is a constant that is sensitive largely to the surface layers, and $D$ depends upon the sound-speed
gradient in the deep interior of the star.

A near-regular pattern of modes in the power spectrum will produce
significant peaks in the PS$\otimes$PS at locations corresponding to
the prominent frequency separations. The most and next-most
significant peaks are usually those corresponding to the separations
$\Delta\nu/2$ and $\Delta\nu/4$, respectively (the former corresponding
to the average separation between modes of odd and even degree,
$\ell$).
Before the PS$\otimes$PS is computed the power spectrum is first
corrected for the background, by division of the best-fitting
background model given by the procedure described in
Section~\ref{sec:bg} above. The PS$\otimes$PS spectrum is then
calculated from a window of the power spectrum centred on $\nu_{\rm
  max}$, as per the MWPS method. For our tests here we set the
frequency window in power spectrum in advance, centered on the true
location of $\nu_{\rm max}$ (rather than run a moving window through
the spectrum).

The statistics of the individual and independent bins in the
PS$\otimes$PS are also described to excellent approximation by $\chi^2$
2-d.o.f. distribution. The probability of observing a relative power in
a single bin greater than or equal to $s$ (i.e. the p-value, with $s$
the height divided by the background level of the PS$\otimes$PS) is
just:
 \begin{equation}
 p(s) = \exp(-s).
 \end{equation}
We may then determine the probability of observing a single bin with a
value of $s$ by chance out of $N$ bins of the PS$\otimes$PS by
 \begin{equation}
 P_{\rm chance} = [1-p(s)]^{\xi N},
 \end{equation} 
where $\xi=1$ for a non-oversampled PS$\otimes$PS and $\xi
\simeq3$ for an oversampling by a factor of 10 (Chaplin et al. 2002).
The final probability of observing a value greater than or equal to
$s$ amongst $N$ bins in the PS$\otimes$PS spectrum (i.e., \emph{not}
by chance) is\footnote{An alternative approach would be to calculate
  the joint probability of observing the three highest values in the
  intervals around $\Delta\nu/2$, $\Delta\nu/4$ and $\Delta\nu/6$ (see
  Hekker et al. 2010). One could also calculate the probability of
  observing the sum of points in the interval around the assumed
  $\Delta\nu/2$, much as the approach followed in the MWPS method}:
 \begin{equation}
 p_{\rm value} = 1- [1-p(s)]^{\xi N}.
 \label{pval2}
 \end{equation}
Again, if the value of $p_{\rm value}$ falls below $\alpha=0.01$ the case
is flagged as a possible detection of stellar oscillations.
\begin{figure*}
  \centering
  \subfloat[SNR curves.]{\includegraphics[trim=0cm 0.25cm 1cm 0cm,clip=true,width=1\columnwidth]{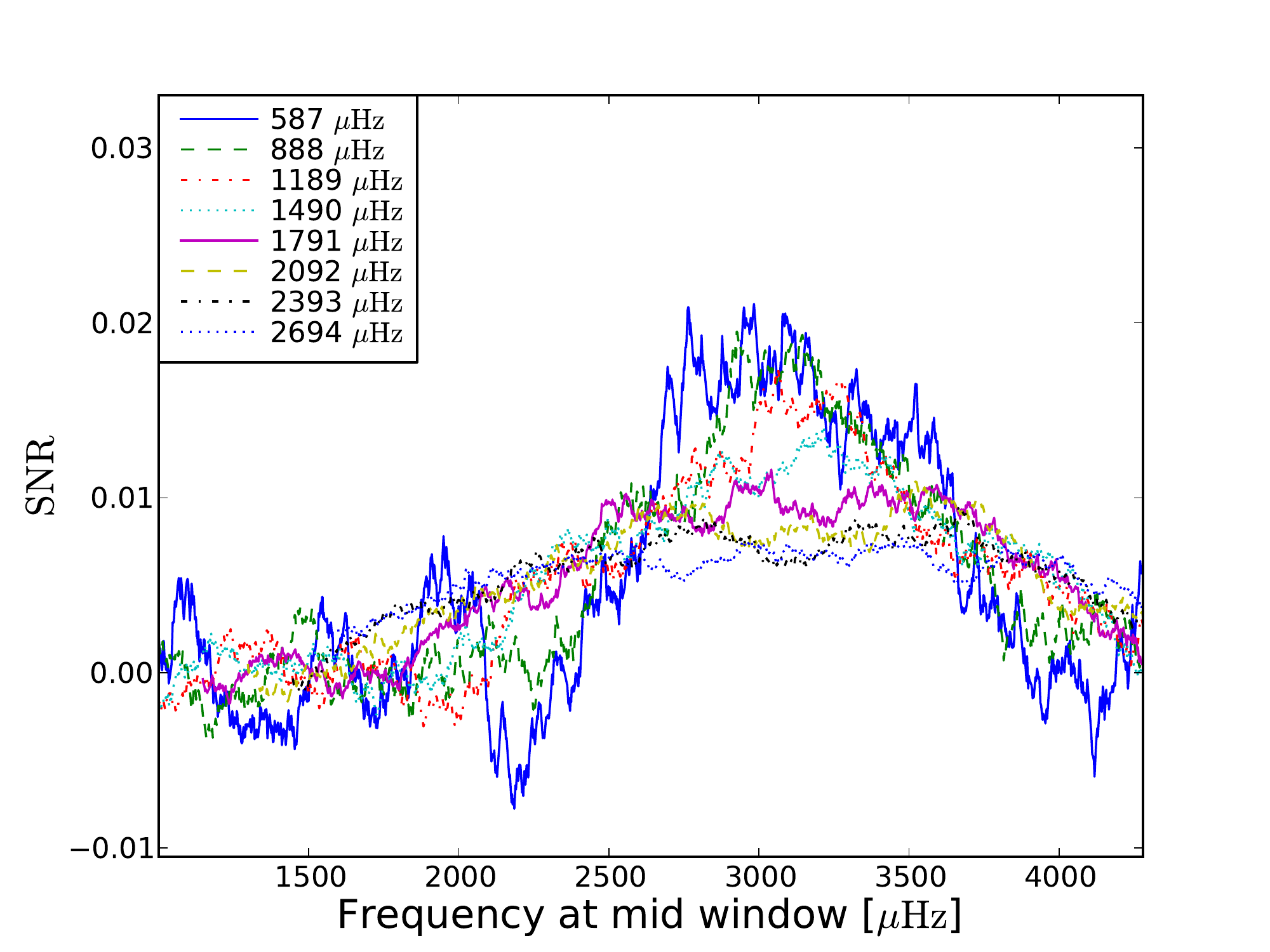}}               
  \subfloat[$1/p_{\rm value}$ curves.]{\includegraphics[trim=0cm 0.25cm 1cm 0cm,clip=true,width=1\columnwidth]{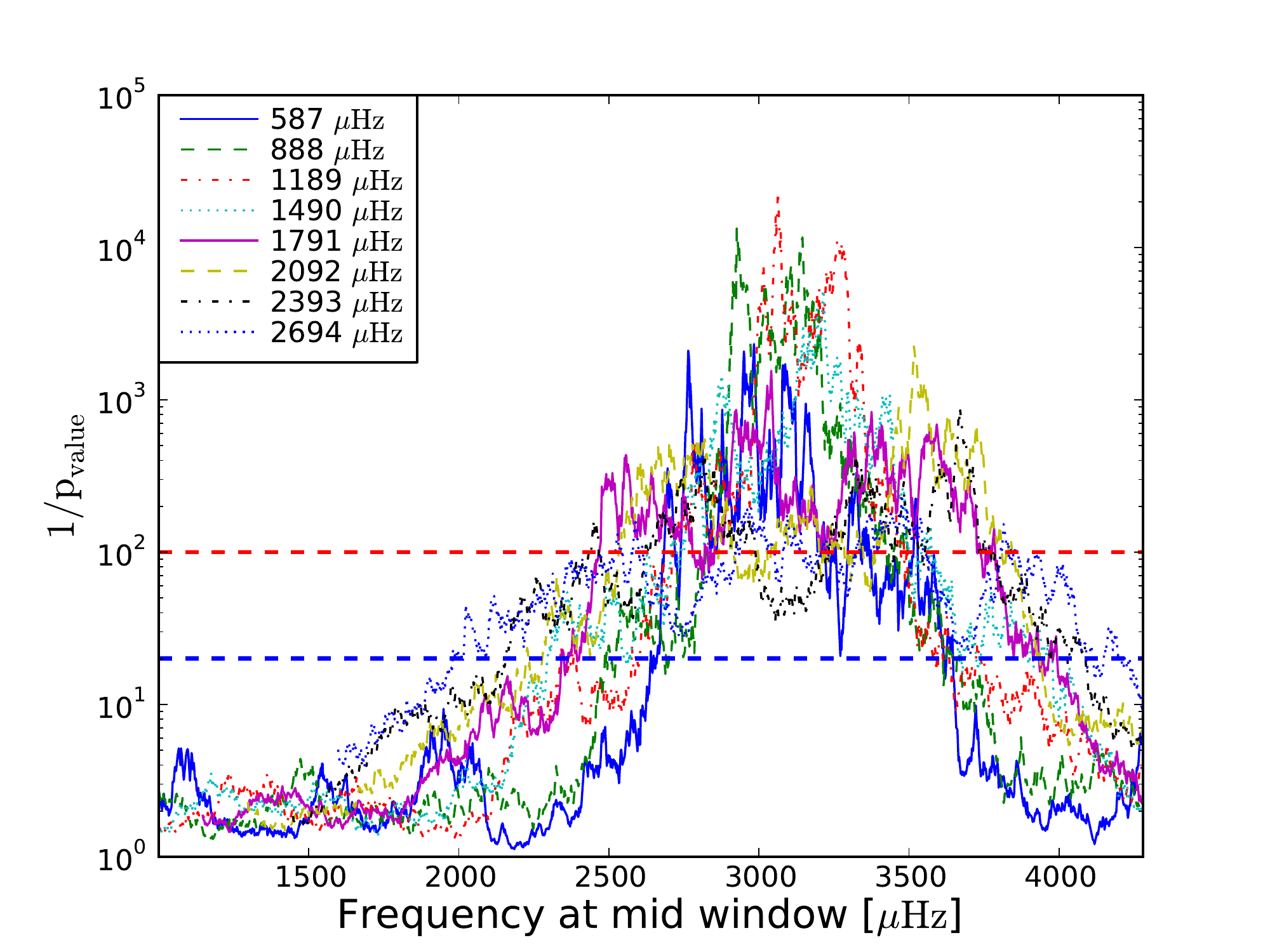}}  
  \caption{A typical result from an MWPS run, here for a realisation
    with $\sigma~=~280$ ppm noise case. Panel (a) shows SNR curves for
    a representative sample of the 40 window widths (with widths
    indicated on the legend). Panel (b) shows the corresponding
    $1/p_{\rm value}$ curves. The red dashed line marks the one per
    cent significance level while the dashed blue line marks the five
    per cent significance level.}
  \label{fig:p_val2}
\end{figure*}   
It is clear from \eqref{pval2} that the number of bins $N$ used in the
computation is crucial in determining the absolute value of $p_{\rm
  value}$. Taking $N$ to be the total number of bins in the
PS$\otimes$PS would only yield detections in very high SNR cases.
However, we may use \emph{a priori} knowledge on stellar oscillations
to restrict the adopted value of $N$. The value of $\Delta\nu$ scales
with $\nu_{\max}$ (Stello et al. 2009) according to
 \begin{equation}
 \Delta\nu = \alpha(\rm \nu_{max}/\mu Hz)^{\beta}
 \label{eq:scaling}
 \end{equation}
with values of $\alpha = 0.22$ and $\beta = 0.797$ (Huber et
al. 2011).  Predictions from stellar evolutionary models suggest that
the value of $\Delta\nu$ given by the relation is accurate to around
15\,per cent, which is the uncertainty we therefore adopt to define
the search range around the $\Delta\nu/2$ value (having adopted the
centre of the frequency window in the power spectrum as the proxy for
$\nu_{\rm max}$). This decreases the value of $N$ significantly,
compared to the total number of bins in the PS$\otimes$PS, and ensures
that the method performs as effectively as possible when compared to
the MWPS method.

\section{\textbf{MWPS and PS$\otimes$PS Tests}}
\label{sec:test}

Both methods have been tested using a 500-day-long time series of
photometric data on the Sun, obtained in the green band (500nm) of the
VIRGO-SPM instrument (a full-disc Sun photometer) on board the
ESA/NASA \textit{SOHO} spacecraft. This time series has a duty cycle
of nearly 99 per cent (Roca Cort\`{e}s et al. 1999), which, combined
with its long duration, makes it safe to assume independent frequency
bins in the power spectrum with $\chi^2_2$ statistics. To this
timeseries were added different levels of random Gaussian noise, with
sample standard deviations covering the range
$\sigma~=~\{180\ \text{ppm},200\ \text{ppm},\ldots,280\ \text{ppm}\}$,
resulting in a total of six tested SNR levels. For each noise level we
generated 20 power spectra, each having a different realization of the
added random noise.

For all the power spectra we omitted the first $250\rm\ \mu Hz$ in the
fit of the background, and found sufficiently good fits (in the sense
of a low $\chi^2$-value) when including only the granulation and shot
noise components. The chosen lag in frequency of the MWPS was $\sim
2.32 \rm\ \mu Hz$ (equivalent to 100 frequency bins), which gives very
good resolution. The bin size for the background binning was chosen as
$\Delta_{\rm bin}=0.05$ (linear in $\text{ln}(\rm \nu)$). The envelope
width of power excess was for all $\nu_{\rm max}$ proxies was taken to
be $0.59$-times $\nu_{\rm max}$.

An example of one of the background fits is plotted in
Fig.~\ref{fig:DJ}. It shows results on one of the realisations with
$\sigma = 280$ ppm. The plotted power spectrum has been smoothed with
a $1\rm\ \mu Hz$-wide boxcar filter. The red points show the binned
spectrum, to which the background function was fitted. The region with
the longest run detected is shown by the green points, while the blue
points show the additional points removed on the low- and
high-frequency sides of the longest run. We also show the background
fits to the points before (magenta curve) and after (black curve) the
removal of the blue and green points. 
While the change in the background fit is quite subtle and hardly visible in this case, the reduced $\chi^2$ did change favourably, from a value of $\chi^2_r =1.24$ with 66 d.o.f. to $\chi^2_r = 1.06$ with 51 d.o.f.
Fig.~\ref{fig:p_val2} shows the MWPS results for the data plotted in
Fig.~\ref{fig:DJ}.  Panel (a) shows the SNR (i.e., $\rm PBR-1$) curves
for several different window sizes (see plot anotation). 
The window sizes used here were selected as described in Section~\ref{sec:width}.
A value of zero on the ordinate in effect corresponds to the total absence of
oscillation signal. Negative values mean that the background was
estimated to be higher than the actual observed power. Panel (b) shows
the corresponding $1/p_{\rm value}$ curves. The dashed red line marks
the one per cent significance level while the dashed blue line marks
the five per cent significance level.  Note that even though the
most narrow window dominates the SNR curve its $1/p_{\rm value}$ is
not as prominent due to the fact that its window comprises a smaller
number of d.o.f. compared to the other windows.
 
\begin{figure}
\centering
\includegraphics[trim=0.4cm 0.2cm 2cm 1.3cm, clip=true, width=1\columnwidth]{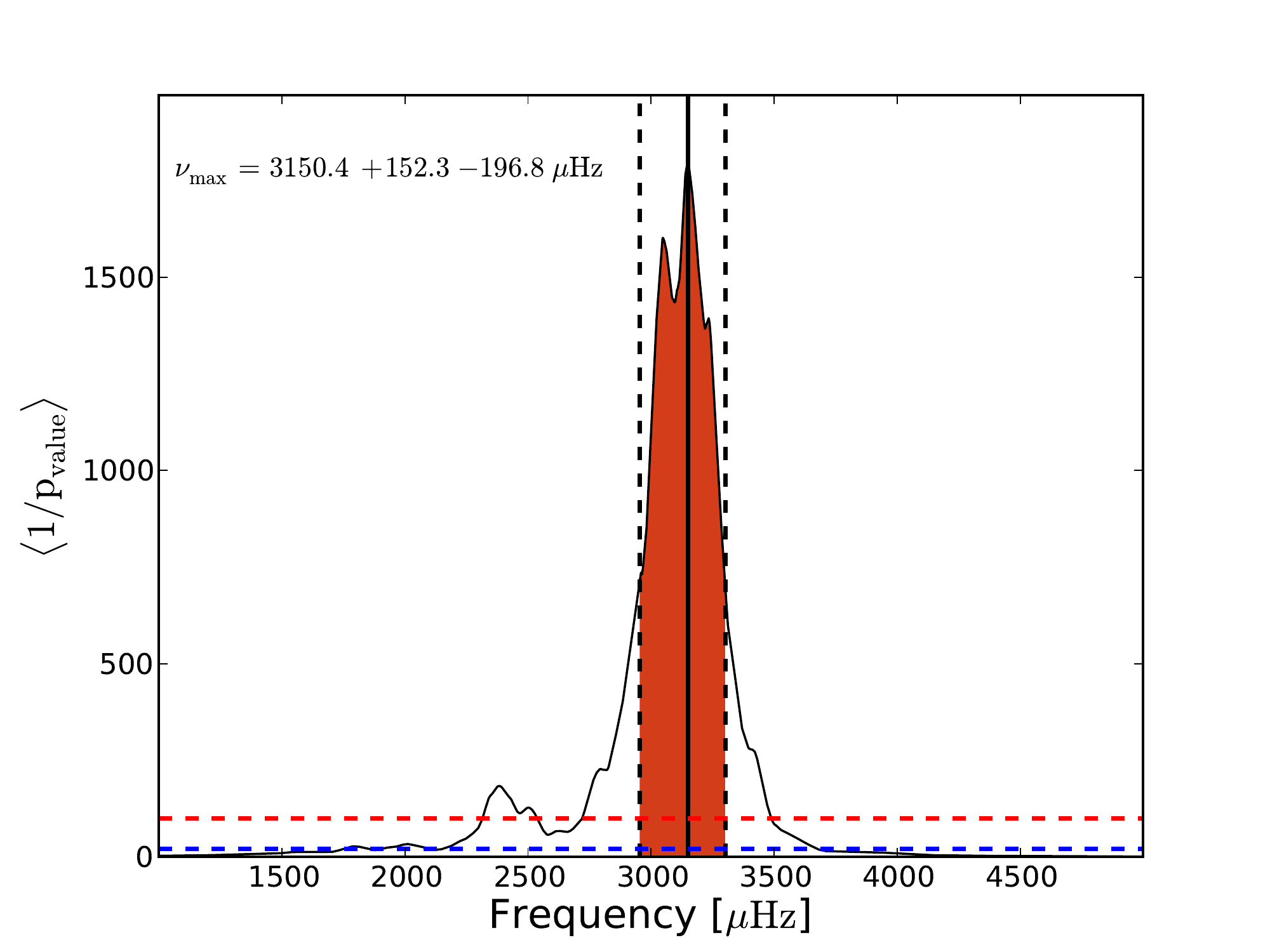}
\footnotesize

\caption{Smoothed average $1/p_{\rm value}$ curve, again for one of
  the $\sigma~=~280$~ppm realizations. The solid black line gives the
  estimated $\nu_{\rm max}$, while the two dashed black lines
  represent the confidence interval (equivalent 1-$\sigma$
  interval). The dashed red and blue lines correspond, respectively, to
  the one and five per cent significance levels.}

\label{fig:average} 
\end{figure}

We estimate $\nu_{\rm max}$ from the $1/p_{\rm value}$ curves as
follows. We begin by calculating a mean $1/p_{\rm value}$ curve from
all curves in which a detection was flagged.  Using the mean curve
instead of just the individual $1/p_{\rm value}$ curves ensures that
if there are spurious peaks (e.g. from very narrow windows) these
will not have a large impact as they are not seen in wider
windows. The mean $1/p_{\rm value}$ curve is then smoothed with a
Savitzky-Golay filter (which ensures that the area under the curve,
the peak positions, and the width/height ratio of peaks, is
conserved), with a window width of 61 points (corresponding to $\rm\sim 142\mu Hz$, with an adopted lag frequency of about $\rm 2.32\mu Hz$). The estimate of $\nu_{\rm max}$ is then given by the
frequency at the highest value in this smoothed curve. To estimate the
confidence limits on the estimated $\nu_{\rm max}$ we treat the
smoothed, mean $1/p_{\rm value}$ curve as a probability density
curve. Here the confidence limits are found by taking the 68.27$\%$ confidence intervals around $\nu_{\rm max}$ (see Barlow
1989). This is calculated by finding the shortest interval around
$\nu_{\rm max}$ wherein 68.27$\%$ of the area under the curve is
contained, equivalent to a one sigma limit for a Gaussian probability
distribution. An example of an average $1/p_{\rm value}$ curve with
estimated confidence limits can be seen in Fig.~\ref{fig:average}.

For the computation of the significance in the PS$\otimes$PS we used the same background fits for correcting the power spectrum as were used for the MWPS method. For the search range, the adopted variation in the value of $\Delta\nu/2$ of $\pm15$ per cent gave $N=13$ bins to be used in \eqref{pval2}. The background level in the PS$\otimes$PS spectrum (not the power spectrum) was found by first smoothing the PS$\otimes$PS spectrum with a median filter. A low-order polynomial was then fitted to the smoothed spectrum, where the search range around $\Delta\nu/2$ was excluded from the fit. The PS$\otimes$PS was then divided by the fitted polynomial curve, giving relative power in the PS$\otimes$PS spectrum.

\begin{figure*}
  \centering
  \subfloat[Non-oversampled PS$\otimes$PS spectrum, $\sigma=180$ ppm noise.]{\includegraphics[width=1\columnwidth]{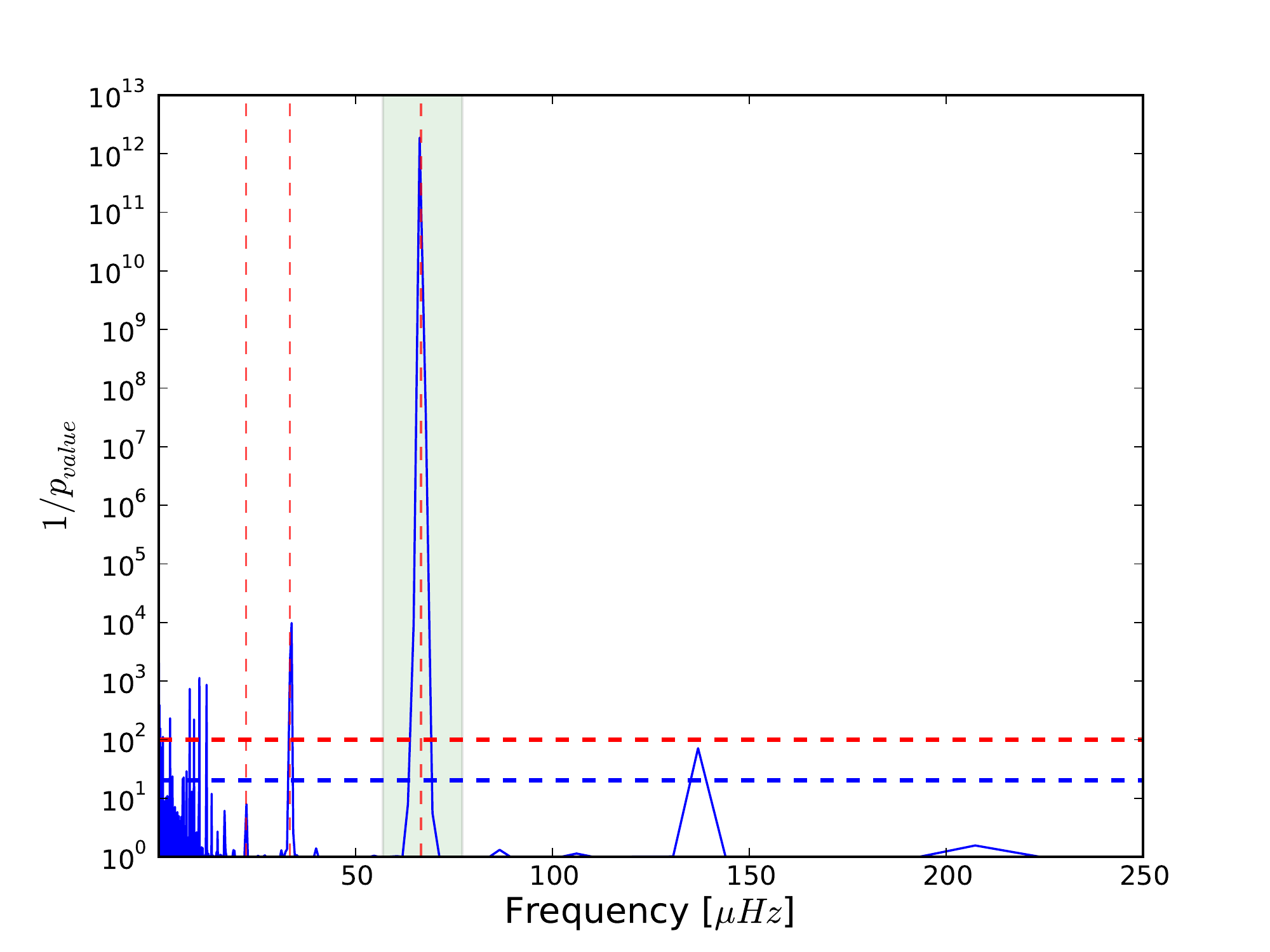}}                
  \subfloat[PS$\otimes$PS spectrum oversampled by a factor of 10, $\sigma=180$ ppm noise.]{\includegraphics[width=1\columnwidth]{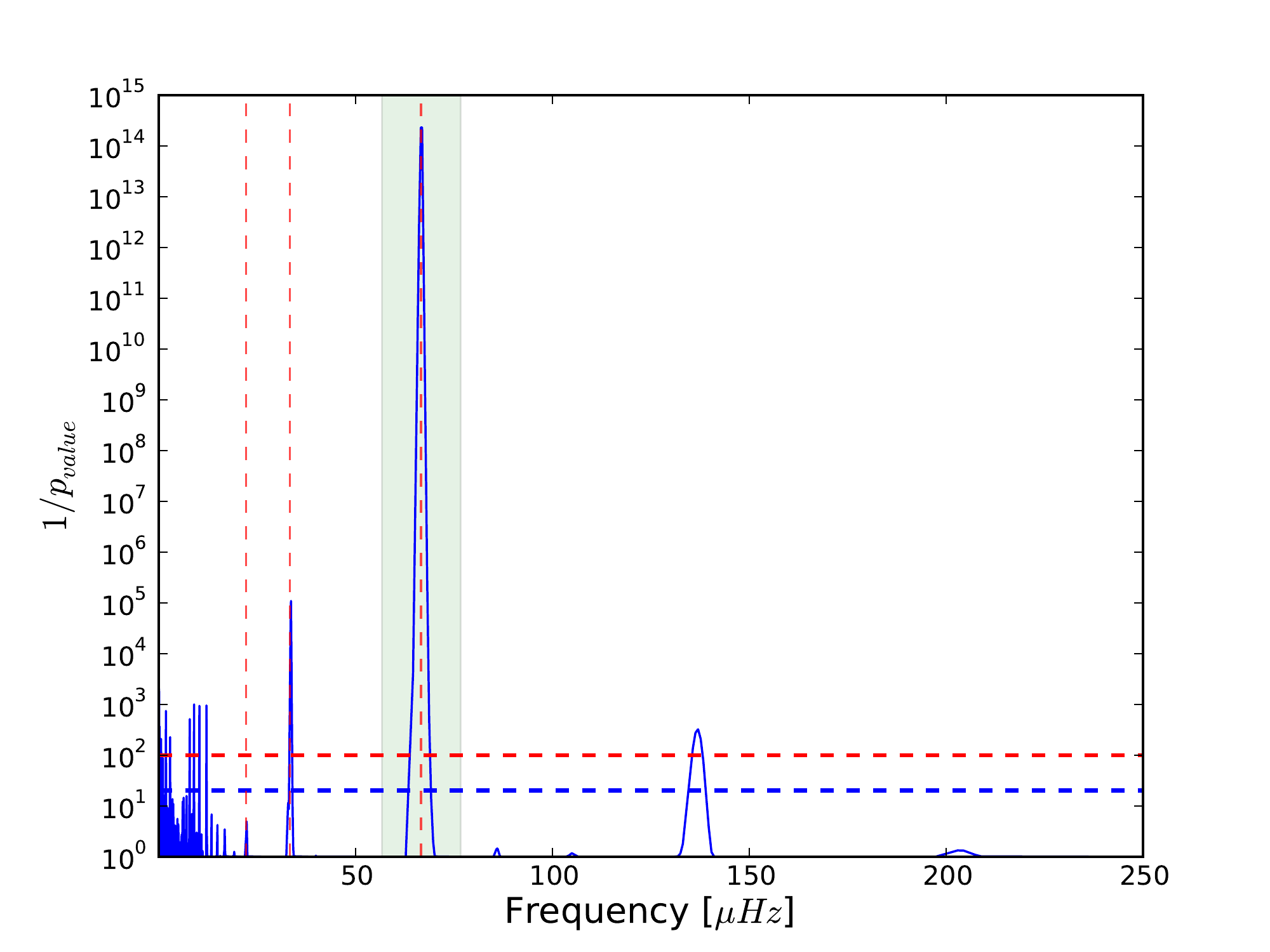}}\\
  \subfloat[Non-oversampled PS$\otimes$PS spectrum, $\sigma=280$ ppm noise.]{\includegraphics[width=1\columnwidth]{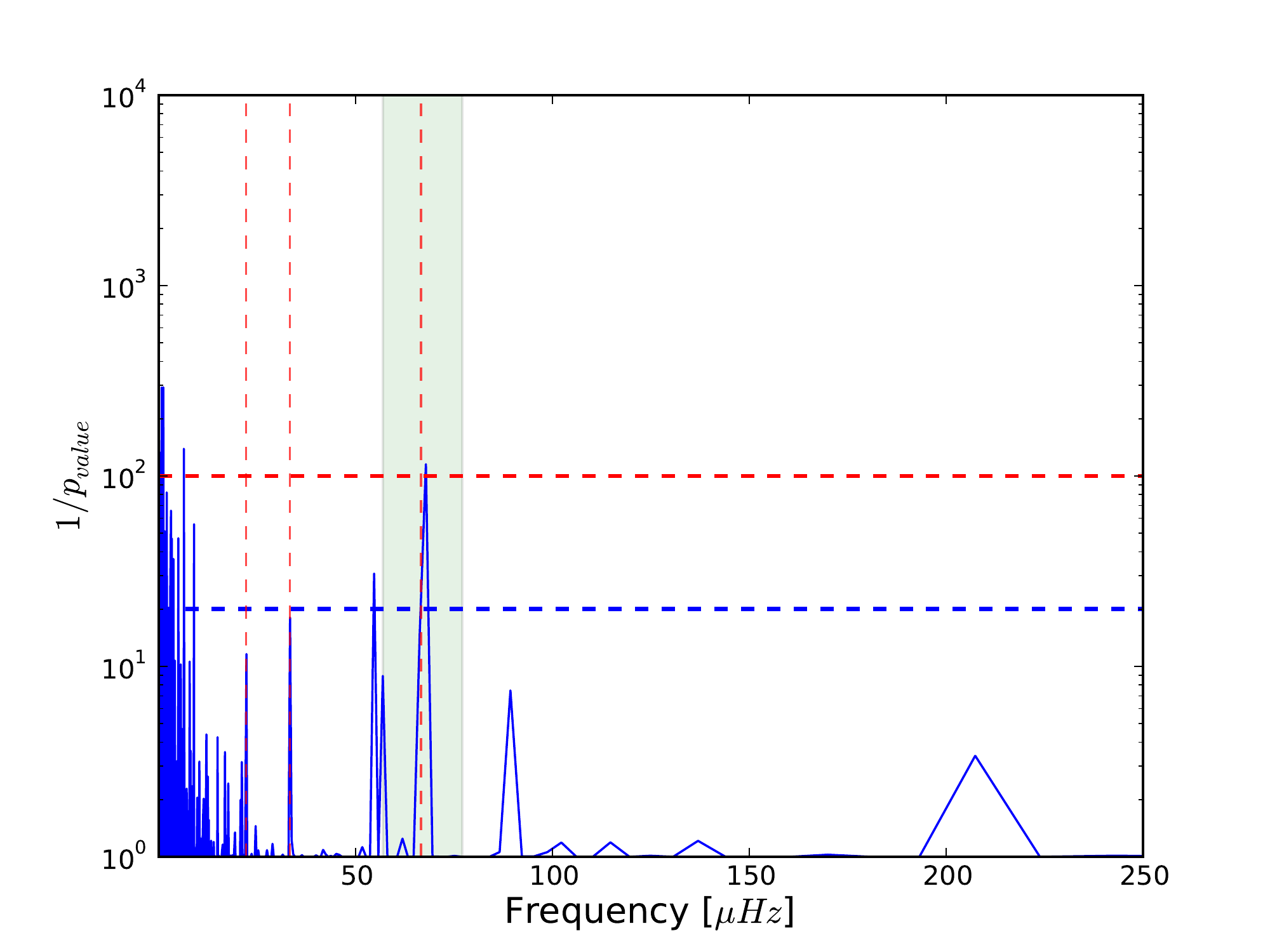}}
  \subfloat[PS$\otimes$PS spectrum oversampled by a factor of 10, $\sigma=280$ ppm noise.]{\includegraphics[width=1\columnwidth]{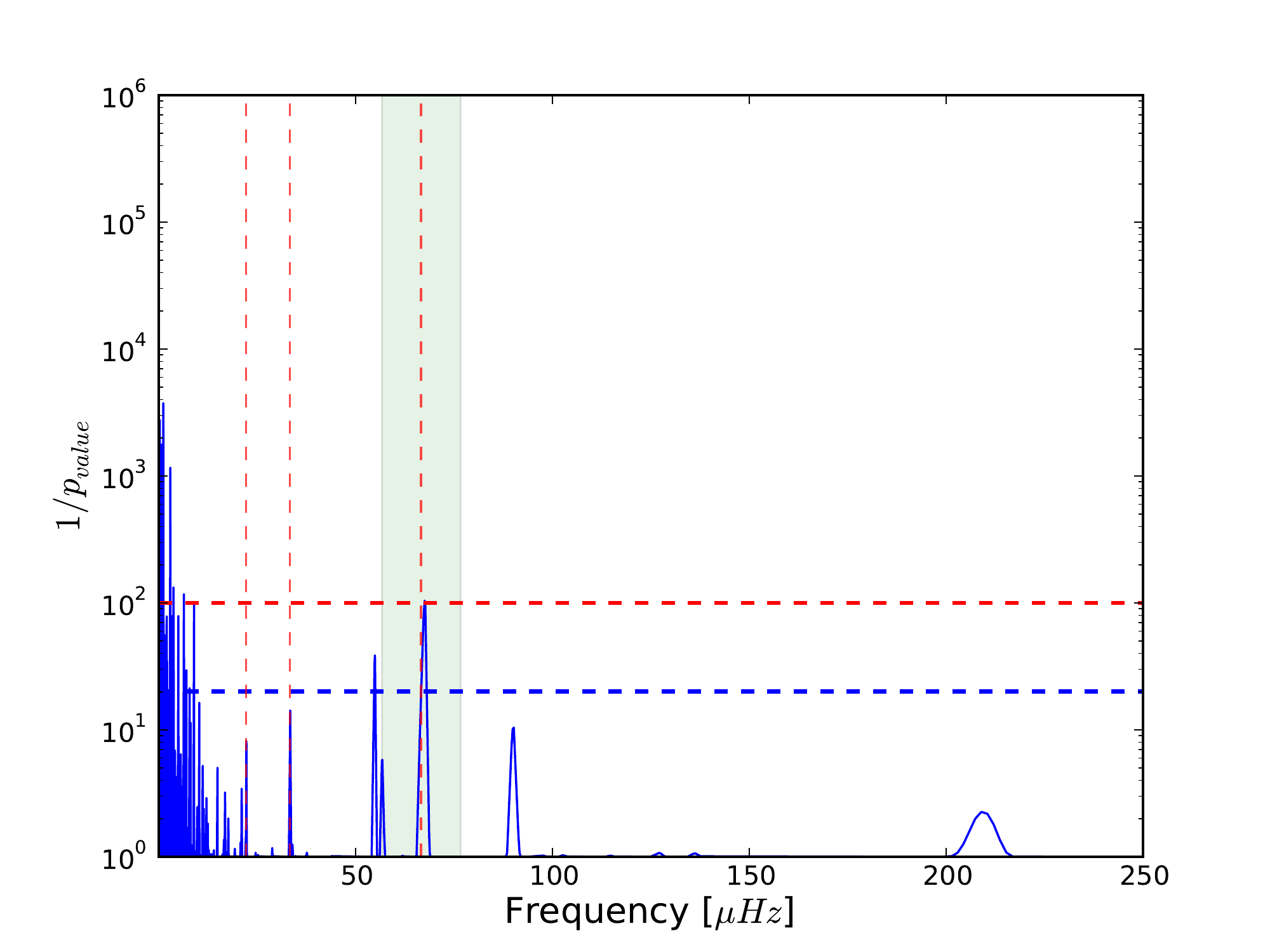}}
  
  \caption{Results of applying the PS$\otimes$PS method to one of the
    $\sigma=180$ and $\sigma=280$ ppm noise realizations. The spectra shown in this figure, each have a $1/p_{\rm value}$ that lies close to the median value for the respective noise levels, see Fig.~\ref{fig:comp2}. Panel (a): $1/p_{\rm value}$ for non-oversampled
    PS$\otimes$PS spectrum of the background corrected power
    spectrum. Here, the PS$\otimes$PS spectrum has been divided
    through by a smoothed background curve, and the $1/p_{\rm value}$ has been calculated using Eq.~\ref{pval2}. The shaded area shows the $\pm15$ per cent
    interval around $\Delta\nu/2$ which is used as the search range
    for significant power. Panel (b): oversampled result for the same
    dataset. Panels (c) and (d): results of applying the
    PS$\otimes$PS method to one of the $\sigma~=~280$ ppm noise
    realizations (same data used to generate MWPS results in
    Fig.~\ref{fig:p_val2}). The red and blue dashed lines mark the one per cent and five per cent significance levels respectively. The dashed vertical lines give the position, in frequency, of $\Delta\nu/2$, $\Delta\nu/4$, and $\Delta\nu/6$.}

  \label{fig:psps}
\end{figure*}

Fig.~\ref{fig:psps} shows non-oversampled and oversampled
PS$\otimes$PS results for $\sigma=180$ and $\sigma=280$\,ppm noise
cases (the latter for the same data as Figs.~\ref{fig:DJ},
\ref{fig:p_val2} and~\ref{fig:average}). The coloured regions in the
figure show the range tested for significant power. Results are shown
at both noise levels to illustrate the degradation of the
PS$\otimes$PS signal in going from the lowest to the highest noise
case considered.  The $\sigma=280$ ppm signal may be compared with the
firm detection found at this noise level by the MWPS method.

\subsection{\textbf{Results}} 
\label{result}


\begin{figure*}
\centering
\includegraphics[width=1\textwidth]{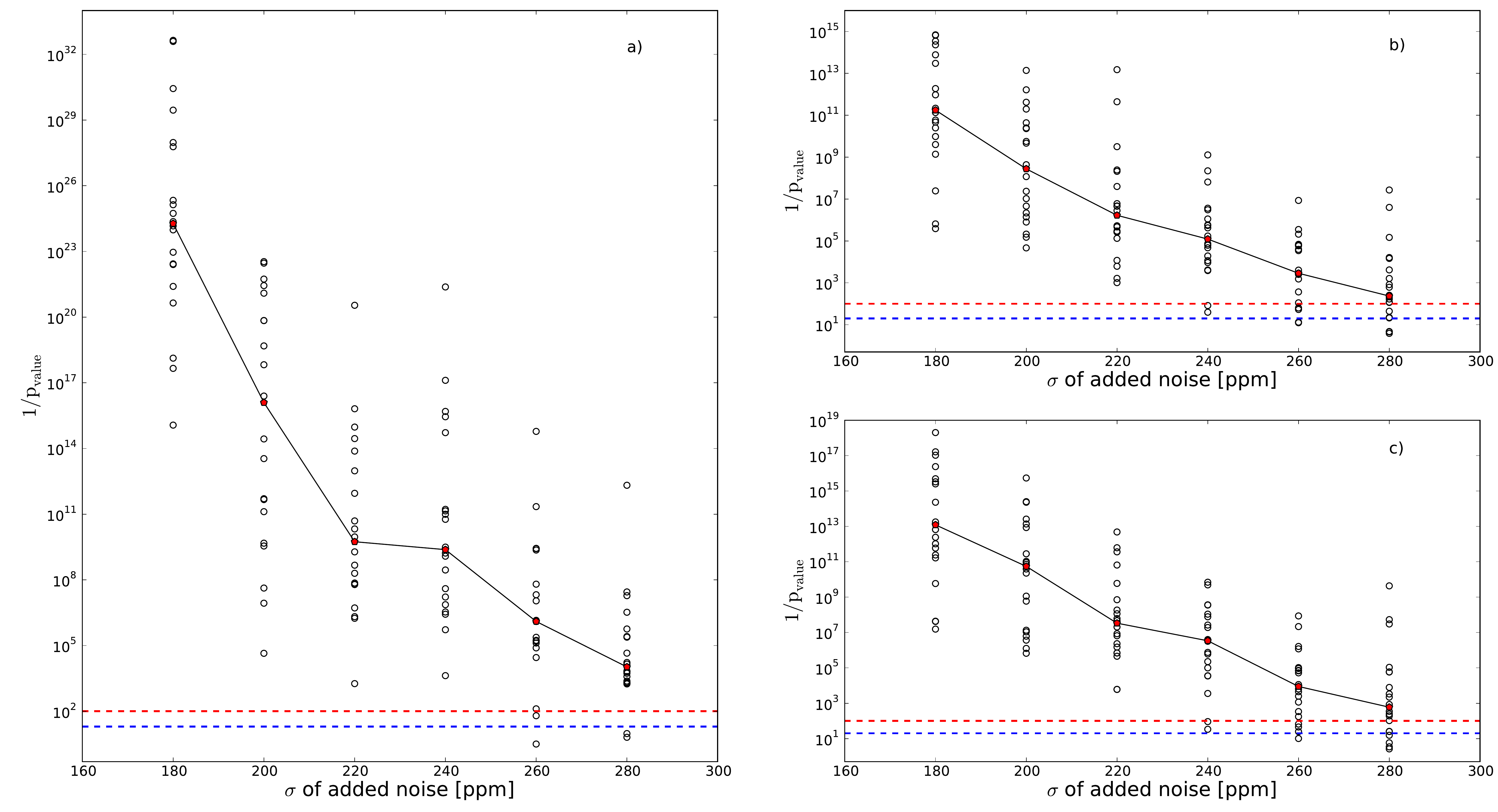}
\footnotesize
\caption{Results from the 20 different realisations of the $\sigma=180$ ppm to $\sigma=280$ ppm noise cases. Panel (a) shows the results obtained from the MWPS method, while panels (b) and (c) show the corresponding results from the PS$\otimes$PS method (no oversampling, and over-sampling by a factor of 10). The values in the MWPS plot, given by open symbols, are the maxima $1/p_{\rm value}$ of the smoothed, average $1/p_{\rm value}$ curves for the individual realisations. In all panels the open symbols show results of the individual realisations, while the red symbols are the median $1/p_{\rm value}$ value over all realisations at each noise level. The red and the blue dashed lines show the threshold values for, respectively, a one per cent and a five per cent detection threshold.}

\label{fig:comp2} 
\end{figure*}



\begin{figure*}
 \centering
 \includegraphics[width=1\textwidth]{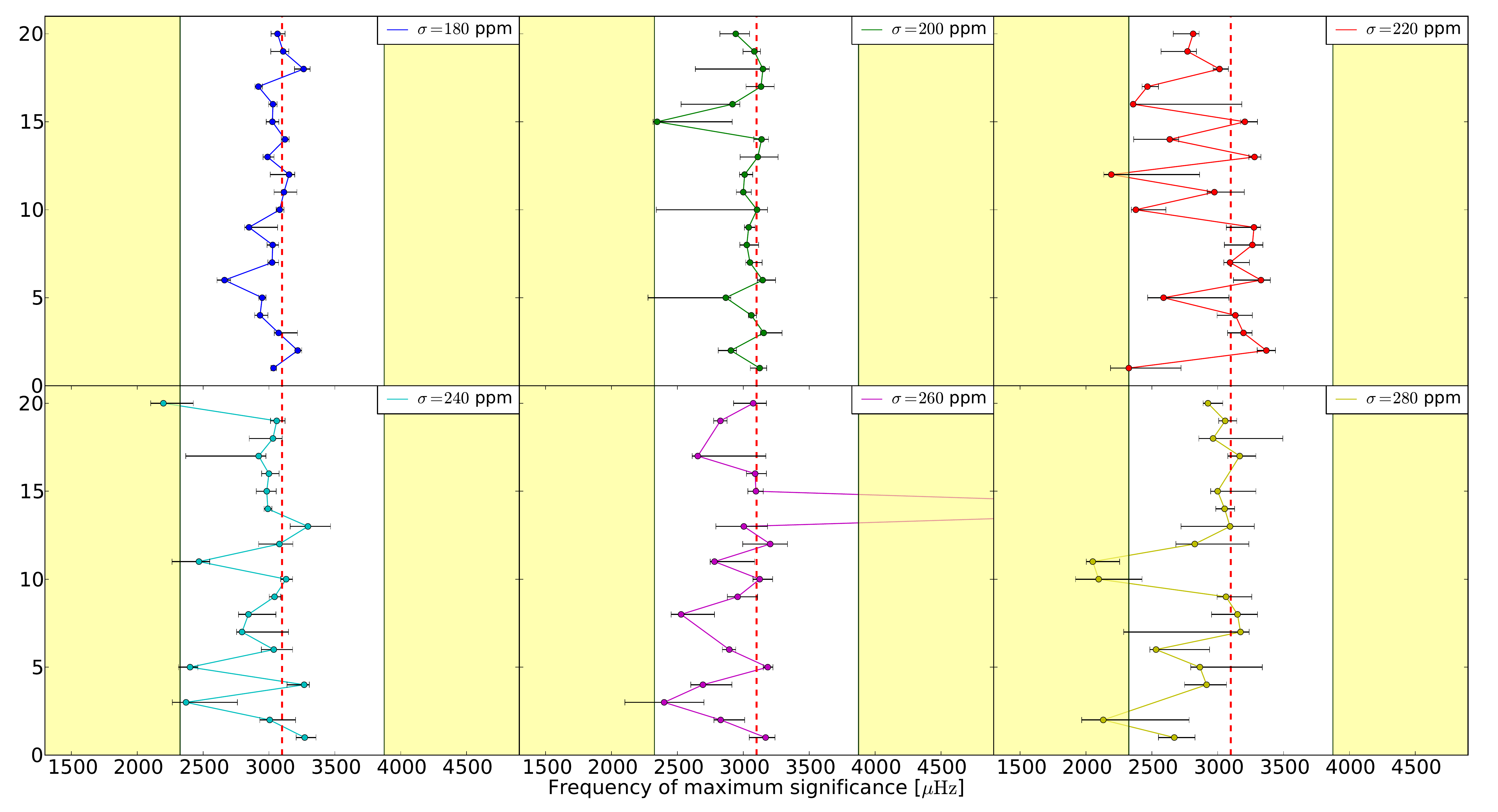}
 \footnotesize
\caption{Estimated $\nu_{\rm max}$ values from the MWPS method, for different realizations of each noise level (see plot legends). Note that for the $\sigma=260$ and $280$ ppm only 18 points are plotted. This is due to the fact that both of these noise levels have two realisations which fall below detection at the 1\,per cent level, as also indicated in Fig.~\ref{fig:comp2} }
 \label{fig:pos} 
\end{figure*}

Panel (a) of Fig.~\ref{fig:comp2} shows the $1/p_{\rm value}$ results
from the MWPS tests. A detection is flagged as positive if the
$1/p_{\rm value}$ is greater than 100 for $\alpha=0.01$ or 20 for
$\alpha=0.05$, corresponding to the dashed red and blue lines
respectively. For each of the noise levels the individual realisations
are plotted with empty circles, and the median values for each of the
noise realisations are marked by the red pentagons.
The results for the PS$\otimes$PS method are shown in panels (b) and
(c). Panel (b) shows the results from non-oversampled PS$\otimes$PS
spectra, while panel (c) shows the results when oversampling by a
factor 10 was used. 




It is apparent that the MWPS returns a higher detection rate than the
PS$\otimes$PS method, i.e., not only are the median detection rates
higher, but there are also a much smaller fraction of
non-detections. The MWPS also shows a much larger spread in the
inferred $1/p_{\rm values}$, but as the values found in general give
positive detections at a high significance level this is not a major
cause for concern.

Since in our test data we know the location in frequency of the
oscillations, we are also in a position to say something about the
occurrence rate of false positive detections. Fig.~\ref{fig:pos} plots
the position in frequency of flagged detections (at the one per cent
level). The noise levels are indicated by the colour coded legend. The
solar value of $\nu_{\rm max\odot} = 3100 \rm\ \mu Hz$ is marked by
the dashed red line.  The lines at 2325 and $3875 \rm\ \mu Hz$ mark
the interval we set for a correct positive detection. This interval
corresponds to a one standard deviation of a Gaussian power envelope
having $\nu_{\rm max} = 3100 \rm\ \mu Hz$.  Points that fall in the
shaded areas, outside this frequency range, might therefore be
considered as putative false-positive detections. Only six
realisations over the entire ensemble fell well within the shaded
areas, three of which were from the highest noise case. Except for the
false-positive in the $\sigma = 260$ ppm case we find that in the
remainder of the false-positive cases the MWPS is still picking up
part of the underlying signal, but the central frequencies are offset
due to beating with the noise. These cases could rightfully be seen as noise-affected detections rather than genuine false-positives. This suggests that even under quite extreme SNR conditions, the MWPS may still be used to ascertain if
excess power is present, but the estimate for the actual $\nu_{\rm
max}$ value should not be trusted in very high noise cases.  

When at each noise level we calculate the difference between the
estimated $\nu_{\rm max}$ and the true, underlying $\nu_{\rm
max\odot}$, in units of the estimated uncertainties, we find that the
fraction of detections that lie within $1\sigma$ of the true value
stay fairly constant, at around 50\,per cent. However, the average
uncertainties do of course increase, reaching about 5\,per cent at a
noise level of $\sigma = 260$. To put this fractional precision in the
context of stellar properties estimation, consider, for example, the
use of $\nu_{\rm max}$ to estimate $\log(g)$, via the scaling relation
$\nu_{\rm max} \propto g T_{\rm eff}^{-0.5}$ (see, e.g., Brown et al. 1991; Kjeldsen \& Bedding 1995). Under the assumption of superior precision
in any complementary estimate of $T_{\rm eff}$, a fractional precision
of 5\,per cent in $\nu_{\rm max}$ equates to a similar fracitonal
precision in $g$, and a precision of about 0.02\,dex in $\log(g)$.

With a fraction of non-detections of only 10\,per cent at $\sigma =
280$ ppm we find (two realisations fall below detection at the 1\,per cent level), as mentioned above, that the MWPS can be utilised at
even higher noise levels to ascertain if excess power is present.

In most of the false-positive cases it is clear that there is an
asymmetry in the estimated uncertainties on the central detection
frequency, with the larger uncertainty pointing towards the true
$\nu_{\rm max}$ value. In the bulk of these cases the asymmetry is
caused by multiple peaks with nearly the same height in the mean
$1/p_{\rm values}$ curve (supporting our contention that beating with
the noise is to blame), where the highest peak happens to be at the
low frequency end. The fact that the uncertainty goes mostly towards
the true $\nu_{\rm max}$ value also supports our claim that part of
the underlying signal is being picked up.

From looking at the results presented in Fig.~\ref{fig:psps}, it is clear that, in the sense of determining an accurate estimate for $\nu_{\rm max}$, the PS$\otimes$PS method will be much better. This can be seen from the fact that the $1/p_{\rm values}$ peaks lie, in frequency, at the position of the $\Delta\nu/2$ value. The application of Eq.~\ref{eq:scaling} would therefore return a value for $\nu_{\rm max}$ very close to the true value (under the assumption that the correct peak is used for the computation).

\section{\textbf{Conclusion}}
\label{sec:conc}

We have demonstrated that the MWPS method provides an efficient way of
detecting excess power from solar-like oscillations under low
signal-to-noise conditions, and that it also outperforms the
frequently applied PS$\otimes$PS method. It is worth noting that while
the efficiency of the PS$\otimes$PS will be affected by departures
from near regularity of the mode spacings in frequency (e.g., in
evolved solar-type stars showing mixed modes, or in stars showing
significant variations of $\Delta\nu$ with frequency), such departures
will not affect the performance of the MWPS method. 

It is clear that the method is well suited as the initial step in
the search for stellar oscillations in low SNR targets, and can
provide useful information and guidelines as to where one should look
more carefully for signatures of the near-regular frequency
separations of the modes.  It should also be evident that the greatest
challenge for this otherwise relatively simple method is that of
obtaining a reliable background fit, since inferences drawn rely on
the implicit assumption that the fitted background is close to the
actual, underlying background. We found that the introduction of the
runs-test provided an extremely useful additional constraint to
assess the robustness of the background fitting.

\vspace{1cm}
\subsection*{\textbf{ACKNOWLEDGEMENTS}}

We would like to thank Andrea Miglio for checking the spread of the
large separation $\Delta\nu$ about the used scaling relation. 

WJC acknowledges the financial support of the UK Science and
Technology Facilities Council (STFC). MNL thanks WJC and his
colleagues for their hostpitality during two stays in Birmingham when
the MWPS analysis was developed.

Funding for the Stellar Astrophysics Centre (SAC) is provided by The Danish
National Research Foundation. The research is supported by the ASTERISK
project (ASTERoseismic Investigations with SONG and Kepler) funded by the
European Research Council (Grant agreement no.: 267864)

%


\clearpage

%
%

\end{document}